\documentclass{elsart}
\journal{Physics Letters B}
\usepackage{amssymb}
\usepackage[german,english]{babel}
\usepackage[intlimits,sumlimits,namelimits]{amsmath}
\usepackage[dvips]{epsfig}
\usepackage[dvips]{color}
\usepackage{amssymb,ifthen,shadow}
\usepackage{fancyhdr}
\usepackage[dvips]{epsfig}

\newcommand{\beq}{\begin{equation}}
\newcommand{\eeq}{\end{equation}}
\newcommand{\beqa}{\begin{eqnarray}}
\newcommand{\eeqa}{\end{eqnarray}}

\begin{document}

\begin{frontmatter}
  
  \title{\bf Influence of $NN$-rescattering effect on the photon asymmetry of
    $d(\vec{\gamma},\pi^-)pp$ reaction}
  
  \author{Eed M.\ Darwish\corauthref{eed}} \corauth[eed]{{\it E-mail
      address:} darwish@kph.uni-mainz.de.}
  
  \address{Physics Department, Faculty of Science, South Valley
    University,\\ Sohag 82524, Egypt}
  
  \date{\today}

\begin{abstract}
  The influence of final-state $NN$-rescattering on the beam asymmetry
  $\Sigma$ for linearly polarized photons in $\pi^-$ photoproduction
  on the deuteron in the energy range from $\pi$-threshold through the
  $\Delta$(1232)-resonance has been investigated. Numerical results
  for this spin observable are predicted and compared with recent
  experimental data from the LEGS Spin collaboration. Final-state 
  $NN$-rescattering is found to be quite important and leads to a 
  better agreement with existing experimental data. Furthermore, the
  differences with other theoretical models have been discussed.
  
  \vspace{0.2cm}

\end{abstract}
\end{frontmatter}
%%%%%%%%%%%%%%%%%%%%%%%%%%%%%%%%%%%%%%%%%%%%%%%%%%%%%%%%%%%%%%%%
\section{Introduction}\label{sec1}
%%%%%%%%%%%%%%%%%%%%%%%%%%%%%%%%%%%%%%%%%%%%%%%%%%%%%%%%%%%%%%%%
It is a well known fact that polarization observables allow a further
and much more detailed analysis of the process under study compared to
the differential cross section alone. Because polarization observables
contain a much richer information on the dynamics of the system than
attainable without beam and/or target polarization and without
polarization analysis of the particles in the final state. The reason
for this is the fact that in contrast to the differential cross
section, which is a sum of the absolute squares of the $t$-matrix
elements, these polarization observables contain interference terms of
the various reaction amplitudes in different combinations and,
therefore, may be more sensitive to small amplitudes and to small
contributions of interesting dynamical effects.

In recent years a great effort, both from theoretical
\cite{Log00,Lee,Log04,Dar04a,Dar04b,Dar04c,Dar04d} and experimental
\cite{Go02,GDH02,Kru03,Bur04} points of view, has been devoted to the
analysis of single-pion photoproduction with polarized beams and/or
polarized targets. In \cite{Log00}, $\pi^-$ photoproduction on the
deuteron has been studied within a diagrammatic approach including
nucleon-nucleon ($NN$) and pion-nucleon ($\pi N$) rescattering in the
final state. Special emphasize was given for the analyzing powers
connected to beam and target polarization, and to polarization of one
of the final protons. First preliminary model calculations for the
photon asymmetry $\Sigma$ of the $\vec\gamma d\to\pi^-pp$ reaction
have been given in the pure impulse approximation (IA) \cite{Lee}. The
comparison between these predictions for $\Sigma$ and the preliminary
experimental data from LEGS Spin collaboration \cite{LEGS} gives a
clear indication that the effects of final-state interaction (FSI) may
be important. The deuteron tensor analyzing powers of the reaction
$\vec d(\gamma,\pi^-)pp$ have been studied in the IA \cite{Log04}
without inclusion of any FSI or two-body exchange current
contributions.  In our previous papers \cite{Dar04a,Dar04b,Dar04c},
various polarization observables in inclusive single-pion
photoproduction on the deuteron using a polarized photon beam and/or
an oriented deuteron target have been investigated in the pure IA
only, i.e., by neglecting any FSI effects and possible two-body
contributions to the production operator. In particular, a complete
survey on all single- and double-polarization observables like beam
and target asymmetries was given. In \cite{Dar04d} the influence of
final-state $NN$-rescattering on the helicity structure of the
inclusive reaction $\vec\gamma\vec d\to\pi^-pp$ has been investigated.
The differential polarized cross-section difference for the parallel
and antiparallel helicity states has been predicted and compared with
recent experimental data from MAMI (Mainz/Pavia) \cite{Pedroni}.  It
has been shown that the effect of $NN$-rescattering is much less
important in the polarized differential cross-section difference than
in the unpolarized one.

The photon asymmetry $\Sigma$ is very sensitive to the internal
mechanisms of the reaction and, therefore, can be a very useful test
to impose constraints on the theoretical models. The work has been
partly motivated by preliminary experimental results, for the
$\vec\gamma d\to\pi^-pp$ channel, with the LEGS Brookhaven National
Laboratory \cite{LEGS} which shows strong and not trivial angular
dependences of this observable. In agreement with these preliminary
data, one can see in \cite{Lee,Dar04c} that the predictions in the
pure IA can hardly provide a reasonable description of the data since
major discrepancies are found. As already noted in \cite{Lee,Dar04c},
the effect of $NN$-rescattering is quite important. This means in
particular that the calculation in the spectator nucleon model can
only be considered as a first step towards a more realistic
description of spin observables.

In this letter we investigate, therefore, the influence of final-state
$NN$ interaction on the photon asymmetry for the reaction
$d(\vec\gamma,\pi^-)pp$ in the energy region from $\pi$-threshold
through the $\Delta$(1232)-resonance. To our knowledge, the influence
of $NN$-FSI effect on this spin observable has never been studied
before. The $\pi N$-rescattering contribution has been considered 
as negligible in the region of the $\Delta$(1232)-resonance 
\cite{Dar03,Lev01} and thus it is not considered in the present work. 
Our main goal is to analyze the recent experimental data from
LEGS \cite{LEGS}.  Furthermore, it was an open question whether the
inclusion of rescattering contributions would lead to a good
description of the available data.

In the next section we will define the photon asymmetry $\Sigma$ in
terms of the transition matrix amplitude. In section \ref{sec3} we
will present and discuss the numerical results of our calculations and
compare them with the experimental data and other predictions. Finally, 
we summarize our conclusions in section \ref{sec4}.
%%%%%%%%%%%%%%%%%%%%%%%%%%%%%%%%%%%%%%%%%%%%%%%%%%%%%%%%%%%%%%
\section{Linear photon asymmetry}
\label{sec2}
%%%%%%%%%%%%%%%%%%%%%%%%%%%%%%%%%%%%%%%%%%%%%%%%%%%%%%%%%%%%%%%
The beam asymmetry $\Sigma$ for linearly polarized photons is defined
in analogy to deuteron photodisintegration \cite{Aren88} writing the
differential cross section for linearly polarized photons and 
unpolarized deuterons in the form
\beqa 
\frac{d\sigma}{d\Omega_{\pi}} (\theta_{\pi},\phi_{\pi}) & = &
\frac{d\sigma_0}{d\Omega_{\pi}} (\theta_{\pi}) ~\Big[ ~1 ~+~ 
P_{\ell}^{\gamma}~ \Sigma (\theta_{\pi}) ~\cos 2\phi_{\pi}~\Big], 
\eeqa
where $d\sigma_0/d\Omega_{\pi}$ denotes the semi-inclusive unpolarized
differential cross section of incoherent pion photoproduction on the
deuteron, where only the final pion is detected without analyzing its
energy \cite{Dar03}, $P_{\ell}^{\gamma}$ is the degree of linearly
polarized photons \cite{Aren88}, $\theta_{\pi}$ and $\phi_{\pi}$
represent the polar and azimuthal pion angles and $\Sigma$ is the
photon asymmetry for linearly polarized photons. Then one has 
\cite{Dar04a,Aren88}
\beqa 
\Sigma~\frac{d\sigma_0}{d\Omega_{\pi}} & = & -
{\mathcal W_{00}}\,, 
\eeqa 
with 
\beqa 
{\mathcal W_{00}} & = &
\frac{1}{2\sqrt{3}}~\sum_{smt,m_{\gamma}m_dm_d^{\prime}}(-)^{1-m
  _d^{\prime}}~C^{1 1 0}_{m_{d} -m_{d}^{\prime} 0}
\nonumber \\ & & \times 
\int_0^{q_{max}}dq \int d\Omega_{p_{NN}}~\rho_s ~(\mathcal
M^{(t\mu )}_{sm,m_{\gamma}m_d})^{\star}~\mathcal
M^{(t\mu)}_{s-m,m_{\gamma}-m_d^{\prime}}\,.
\label{WIM}
\eeqa 
denoting with $C^{j_1 j_2 j}_{m_1 m_2 m}$ a Clebsch-Gordan
coefficient, $m_{\gamma}$ the photon polarization, $m_{d}$ the spin
projection of the deuteron, $s$ and $m$ total spin and its projection 
of the two outgoing nucleons, respectively, $t$ their total isospin,
$\mu$ the isospin projection of the pion, $q_{max}$ the maximum value
of pion momentum, $\Omega_{p_{NN}}$ the solid angle of the relative
momentum $\vec p_{NN}$ of the final $NN$ system and $\rho_s$ the phase
space factor. For further details with respect to the kinematical
variables and quantum numbers we refer to our previous work
\cite{Dar03}.

For the transition ${\mathcal M}$-matrix we include, in this work, 
besides the pure IA, the driving term from $NN$-rescattering, so that
the total transition matrix reads
\begin{eqnarray}
\label{threethree}
{\mathcal M}^{(t\mu)}_{sm m_{\gamma}m_d} & = &
{\mathcal M}_{sm m_{\gamma}m_d}^{(t\mu)~IA} +
{\mathcal M}_{sm m_{\gamma}m_d}^{(t\mu)~NN}\,,
\end{eqnarray}
where the first term represents the transition amplitude in the pure
IA and the second is the corresponding one for $NN$-rescattering in
the final state. Further details with respect to the matrix elements 
are not discussed here and can be found in \cite{Dar03}.
%%%%%%%%%%%%%%%%%%%%%%%%%%%%%%%%%%%%%%%%%%%%
\section{Results and discussion}
\label{sec3}
%%%%%%%%%%%%%%%%%%%%%%%%%%%%%%%%%%%%%%%%%%%%%
Here we present and discuss our results for the photon asymmetry
$\Sigma$ for linearly polarized photons of $\pi^-$ photoproduction on
the deuteron with inclusion of $NN$-rescattering in the final state.
These results are also compared to the preliminary experimental data
of LEGS \cite{LEGS} and the preliminary IA calculations of Lee
\cite{Lee}.  The results presented here are calculated by using the
elementary photoproduction operator of Schmidt {\it et al.}
\cite{ScA96} and the deuteron wave function of Paris potential
\cite{La+81}. For the half-off-shell $NN$-scattering amplitude, the
separable representation \cite{HaP8485} of the realistic Paris
potential has been used. All partial waves with total angular momentum
$J\le 3$ are included.

\begin{figure}[htp]
\begin{center}
  \hspace*{-0.5cm}\includegraphics[scale=0.78]{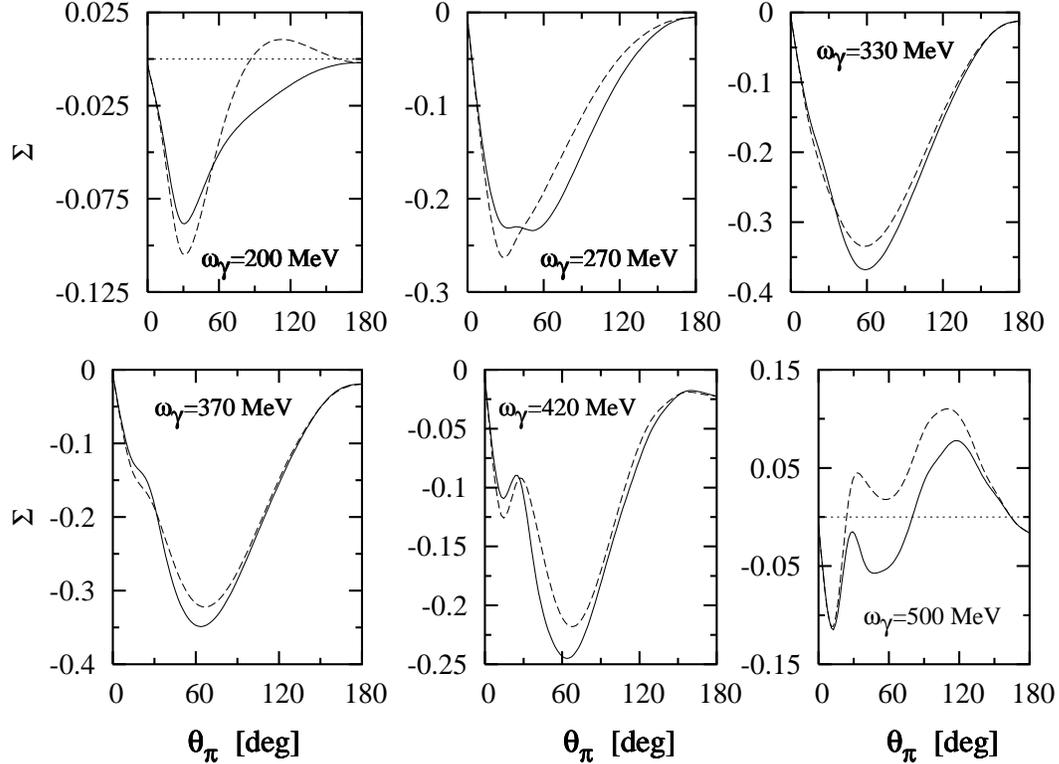}
\caption{Linear photon asymmetry of the differential cross 
  section for linearly polarized photons for $\pi^-$ 
  photoproduction on the deuteron as a function of emission 
  pion angle in the laboratory frame at fixed values of 
  photon lab-energies. Notation of curves: dashed: IA; 
  solid: IA+$NN$-rescattering.}
\label{linear}
\end{center}
\end{figure}

We start with presenting our results for the linear
photon asymmetry $\Sigma$ at different photon lab-energies
$\omega_{\gamma}=200$, $270$, $330$, $370$, $420$ and $500$ MeV in
Fig.\ \ref{linear} as a function of emission pion angle $\theta_{\pi}$
in the laboratory frame. The solid curves show the results of the full
calculation, i.e., when $NN$-rescattering is included, while the
dashed curves show the contribution of the IA alone in order to
clarify the importance of $NN$-FSI effect. 
%%%%%%%%%%%%%%%%%%%%%%%%%%%%%
%%%%%%%%%%%%%%%%%%%%%%%%%%%%%
%%%%%%%%%%%%%%%%%%%%%%%%%%%%%
\begin{figure}[htp]
\begin{center}
  \hspace*{-0.5cm}\includegraphics[scale=0.78]{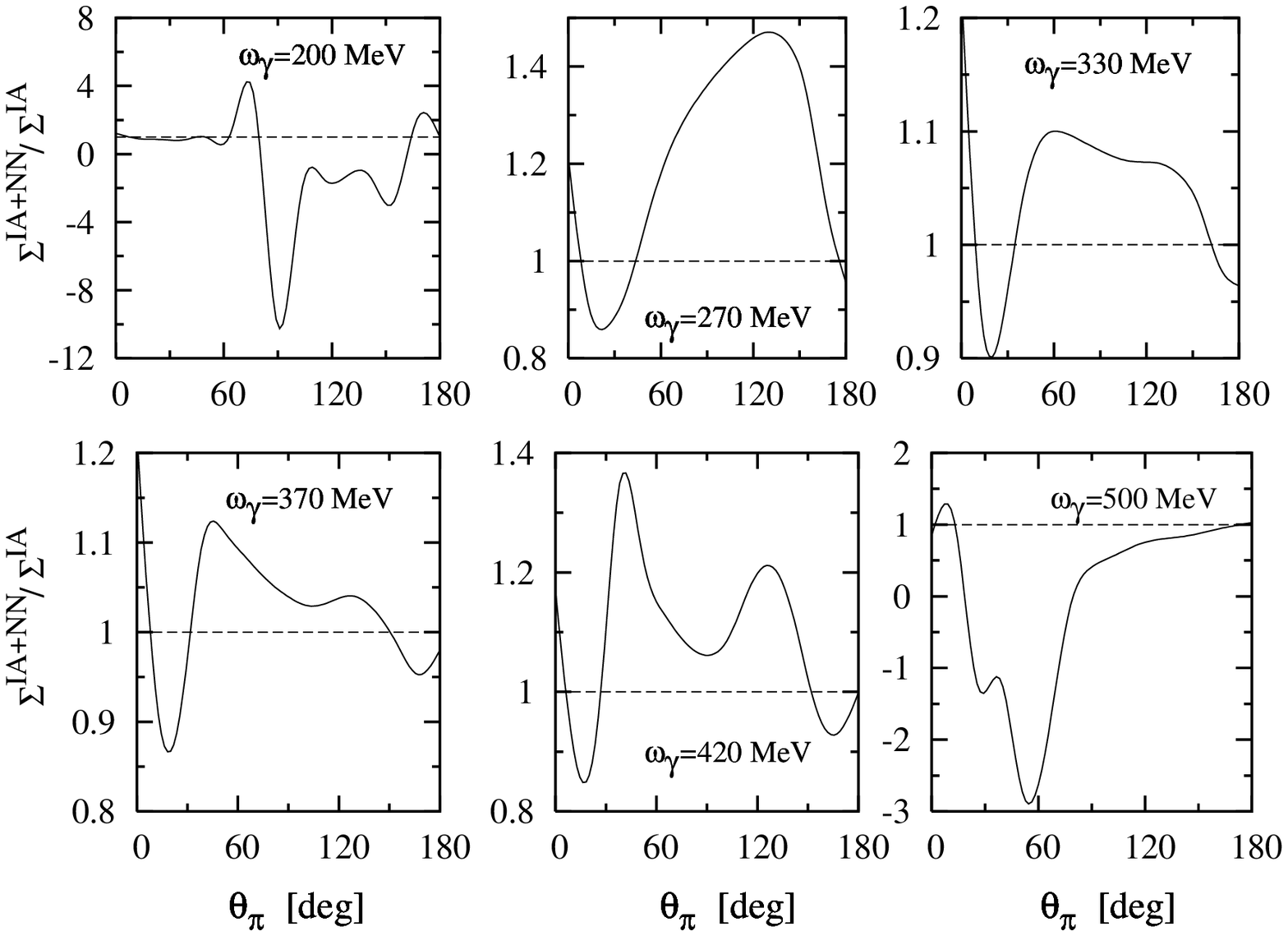}
\caption{The ratio of the linear photon asymmetry with 
  $NN$-rescattering $\Sigma^{IA+NN}$ to the one in the IA
  $\Sigma^{IA}$ as a function of emission pion angle at the same
  photon lab-energies as in Fig.~\ref{linear}.}
\label{ratio}
\end{center}
\end{figure}
%%%%%%%%%%%%%%%%%%%%%%%%%%%%%
In order to show in greater detail the relative influence of
$NN$-rescattering effect on the linear photon asymmetry, we show in
Fig.\ \ref{ratio} the effect of $NN$-rescattering relative to the IA
by the ratio $\Sigma^{IA+NN}/\Sigma^{IA}$, where $\Sigma^{IA}$ denotes
the photon asymmetry in the IA and $\Sigma^{IA+NN}$ the one including
the contribution of $NN$-rescattering.

In the photon energy domain of this work, the magnetic multipoles
dominate over the electric ones, due to the excitation of the
$\Delta$-resonance. This is clear from the dominately negative values
of $\Sigma$ as shown in Fig.\ \ref{linear}. On the contrary, the
left-top and right-bottom panels in Fig.\ \ref{linear} show that small
positive values are found at $\omega_{\gamma}=200$ and $500$ MeV.  We
see also that the asymmetry $\Sigma$ is sensitive to the energy of the
incoming photon. It is noticeable, that the photon asymmetry $\Sigma$
vanish at $\theta_{\pi}=0^{\circ}$ which is not the case at
$180^{\circ}$. At extreme forward and backward emission pion angles
one sees, that the photon asymmetry is relatively small in comparison
to the results when $\theta_{\pi}$ changes from about $30^{\circ}$ to
$120^{\circ}$. One notices also, that the contribution from
$NN$-rescattering is much important in this region, in particular in
the peak position. For lower and higher photon energies, one finds the
strongest effect by $NN$-rescattering.

\begin{figure}[htp]
\begin{center}
  \includegraphics[scale=0.75]{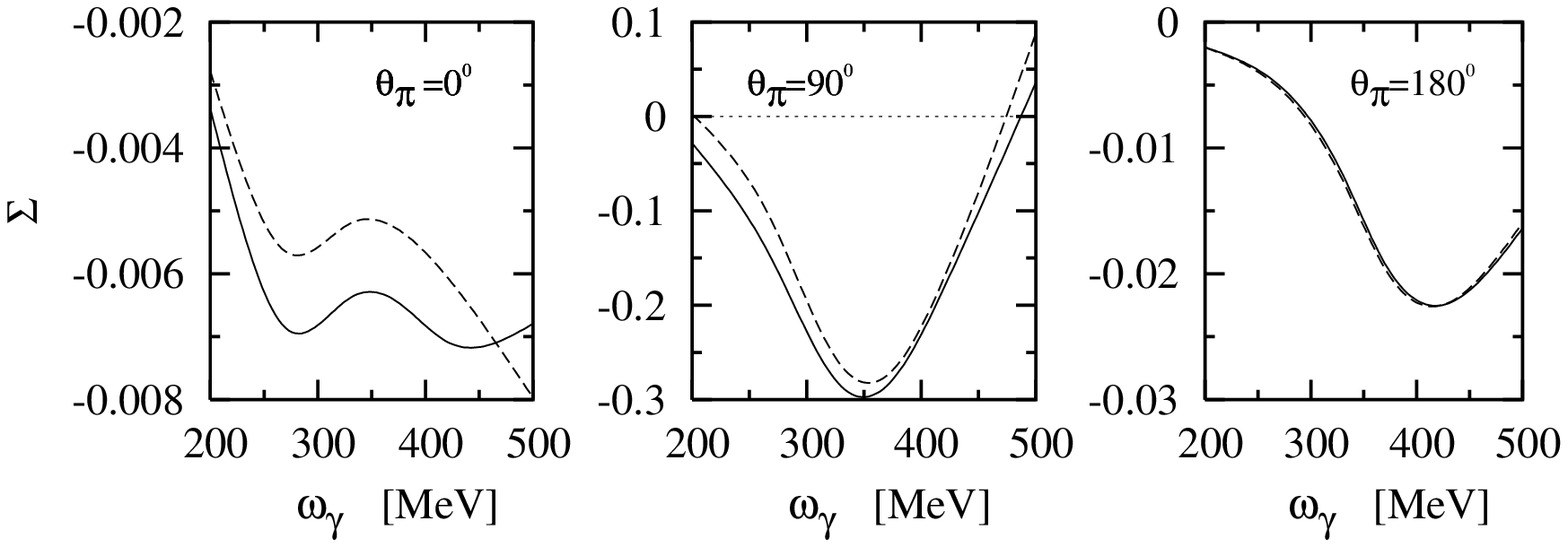}\\
  \includegraphics[scale=0.75]{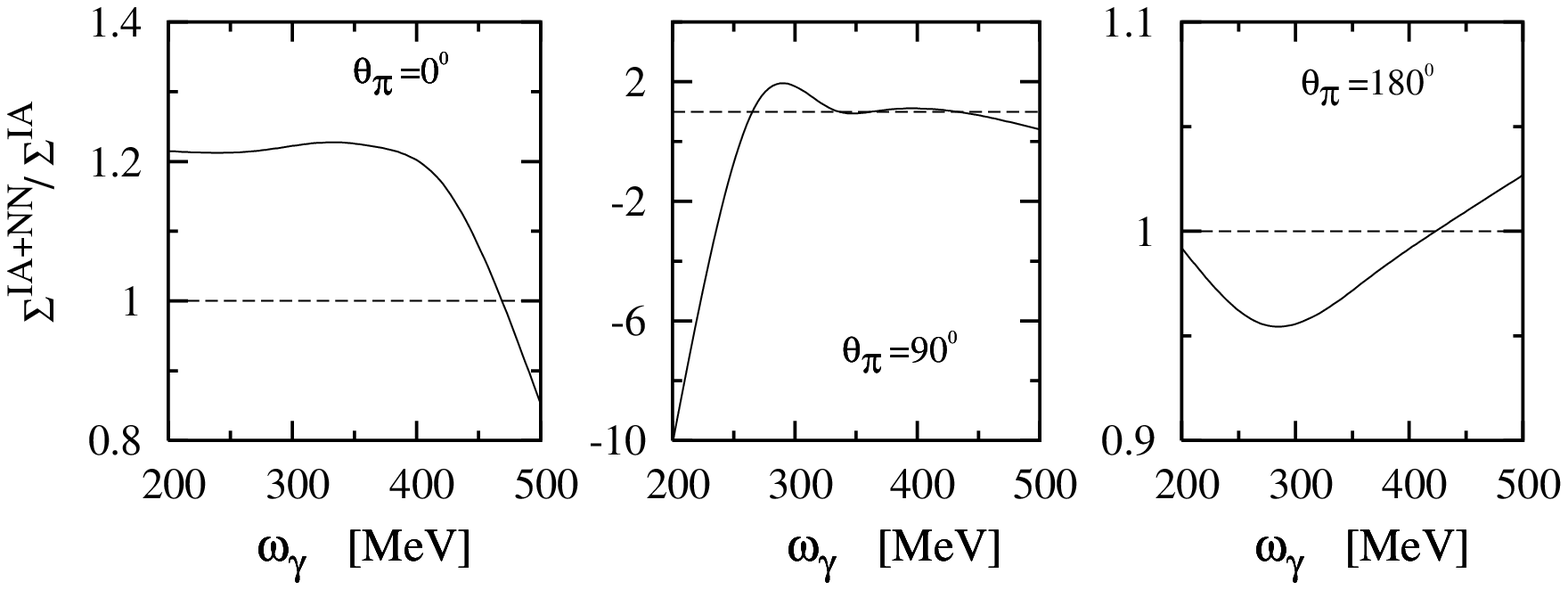}
  \caption{Linear photon asymmetry $\Sigma$ for $d(\vec\gamma,\pi^-)pp$
    as a function of the photon lab-energy $\omega_{\gamma}$ at fixed
  values of pion angle $\theta_{\pi}$ in the laboratory frame with
  $NN$-rescattering (top panels) and their ratios with respect to the
  pure IA (bottom panels). Notation as in Fig.\ \ref{linear}.}
  \label{linearomega}
\end{center}
\end{figure}

Fig.\ \ref{linearomega} shows the sensitivity of our results for the
linear photon asymmetry $\Sigma$ to the photon lab-energy
$\omega_{\gamma}$ at three fixed values of pion angle
$\theta_{\pi}=0^{\circ}$, $90^{\circ}$ and $180^{\circ}$ for photon
lab-energies between 200 and 500 MeV. In order to show the relative
influence of $NN$-rescattering effect on the linear photon asymmetry,
we show in the bottom panels of Fig.\ \ref{ratio} the effect of
$NN$-rescattering relative to the IA by the ratio
$\Sigma^{IA+NN}/\Sigma^{IA}$, where $\Sigma^{IA}$ denotes the photon
asymmetry in the IA and $\Sigma^{IA+NN}$ the one including the
contribution of $NN$-rescattering. In view of these results, one notes
that $NN$-rescattering - the difference between the solid and the
dashed curves - is quite small, almost completely negligible at
extreme forward and backward angles.
%%%%%%%%%%%%%%%%%%%%%%%%%%%%%%
%%%%%%%%%%%%%%%%%%%%%%%%%%%%%%%%%%%

\begin{figure}[htp]
\begin{center}
  \includegraphics[scale=0.75]{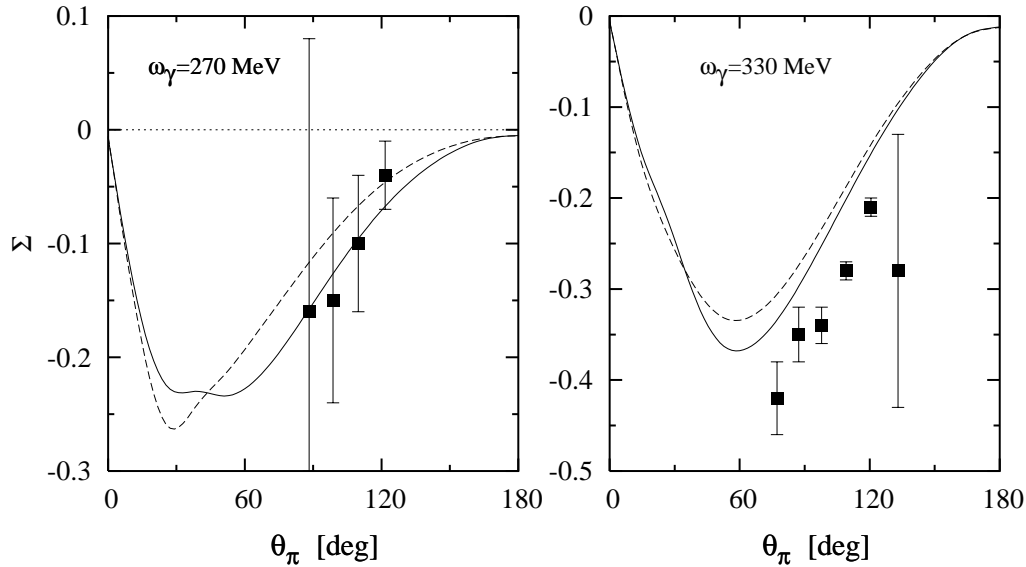}
\caption{Photon asymmetry $\Sigma$ for the reaction 
  $d(\vec\gamma,\pi^-)pp$ at $\omega_{\gamma}=270$ 
  and $330$ MeV photon lab-energy in comparison with 
  the preliminary data from LEGS \cite{LEGS}. 
  Notation as in Fig.~\ref{linear}.}
\label{exp}
\end{center}
\end{figure}

Fig.\ \ref{exp} shows a comparison of our numerical results for the
linear photon asymmetry $\Sigma$ in the pure IA (dashed curves) and
with $NN$-rescattering (solid curves) with experimental data. In view
of the fact that experimental data for this spin observable are not
available in a final form, we compare our predictions with the
preliminary experimental data from the LEGS Spin collaboration
\cite{LEGS} as depicted in Fig.\ \ref{exp}. We see that the general
feature of the data is reproduced. However, the discrepancy is rather
significant in the region where the photon energy close to the
$\Delta$-resonance. This could be due to the higher order
rescattering mechanisms which are neglected in this work. In the same
figure, we also show the results from the IA only (dashed curves). It
is seen that the $NN$-rescattering yields an about 10$\%$ effect in
the region of the peak position. We found that this is mainly due to
the interference between the IA amplitude and the $NN$-FSI amplitude.
In agreement with our previous results \cite{Dar04c}, one notes that
the pure IA (dashed curves in Fig.\ \ref{exp}) cannot describe the
experimental data. The inclusion of $NN$-FSI leads at
$\omega_{\gamma}=270$ MeV to a quite satisfactory description of the
data, whereas at $330$ MeV $NN$-FSI effect is small and
therefore differences between theory and experiment are still evident.
It is appear that our model is still not capable of describing the
measured photon asymmetry, even if $NN$-FSI is included. Future
efforts must be made to remove the remaining discrepancies such as a
complete three-body treatment of the final $\pi NN$ system.
%%%%%%%%%%%%%%%%%%%%%%%%%%%%%
%%%%%%%%%%%%%%%%%%%%%%%%%%%%%

\begin{figure}[htp]
\begin{center}
  \includegraphics[scale=0.75]{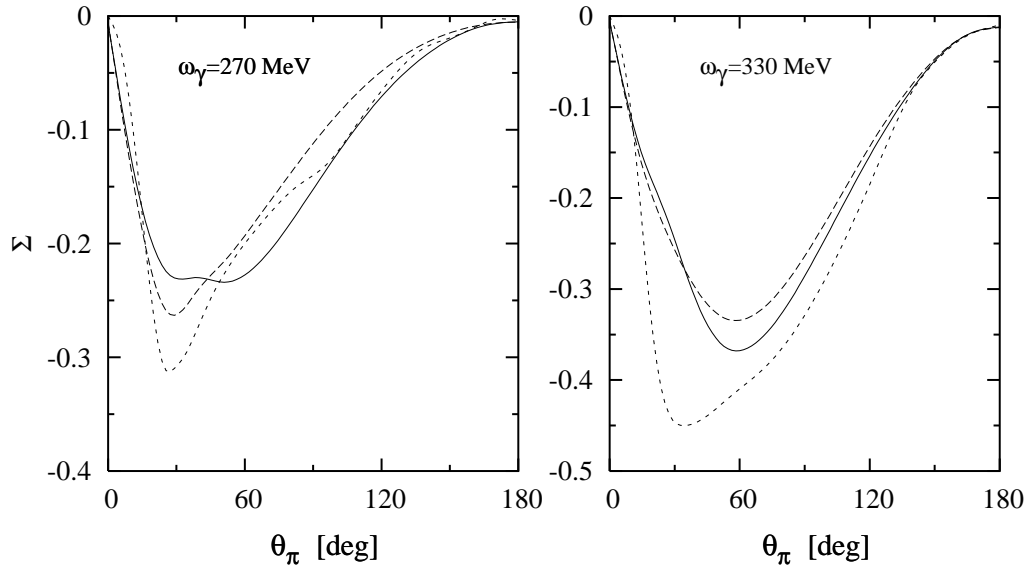}
\caption{Photon polarization asymmetry $\Sigma$ for 
  $d(\vec\gamma,\pi^-)pp$ reaction at two photon lab-energies. 
  Dashed curves: IA of present calculations; solid curves: IA plus 
  $NN$-FSI of present calculations; dotted curves: preliminary IA 
  calculations of Lee \cite{Lee}.}
\label{withlee}
\end{center}
\end{figure}

Now we compare our results for the linear photon asymmetry with the
preliminary model predictions of Lee \cite{Lee} as shown in Fig.\ 
\ref{withlee}. The solid curves show the results of the present
calculations when $NN$-rescattering is included and the dashed ones 
show our results in the IA alone. The preliminary IA results of Lee
\cite{Lee} are represented in this figure by the dotted curves. 
It is very clear that the results for the IA of the present work showed
certain significant differences to the preliminary IA results
\cite{Lee} which cannot be attributed to the use of different
elementary pion photoproduction operators and/or from different
realistic $NN$ potential models used for the deuteron wave function. 
It is clear from the right panel of Fig.\ \ref{withlee} that the
discrepancy is rather significant in the region of the
$\Delta$(1232)-resonance, in particular in the peak position. 

As already mentioned in the beginning of this section, our results are
calculated using the effective Lagrangian model developed by Schmidt
{\it et al.} \cite{ScA96}. The main advantage of this model is that it
has been constructed to give a realistic description of the
$\Delta$-resonance region. It is also given in an arbitrary frame of
reference and allows a well defined off-shell continuation as required
for studying pion production on nuclei. As shown in Figs.\ 1-3 in
\cite{Dar03}, the results for this model are in good agreement with
recent experimental data as well as with other theoretical
predictions. On the other hand, the well-known dynamical model of Sato
and Lee \cite{SL} has been used in \cite{Lee}. This model has given
also a successful description of the pion photoproduction data.
Therefore, the big difference between both predictions in the IA
results cannot be attributed to the use of different elementary
operators. This can be interpreted as lack of understanding of the
nonresonant background, which in dynamical models is related to the
pion cloud. It seems that pion cloud effects are not yet consistently
included in dynamical models. An independent evaluation for $\Sigma$
in the pure IA as well as with the 'totally neglected' rescattering 
mechanisms in \cite{Lee} would be very interesting to understand the 
origin of this discrepancy.
%%%%%%%%%%%%%%%%%%%%%%%%%%%%%
%%%%%%%%%%%%%%%%%%%%%%%%%%%%%%%%%%%%%%%
%%%%%%%%%%%%%%%%%%%%%%%%%%%%%%%%%%%%%%%
\section{Summary}
\label{sec4}
%%%%%%%%%%%%%%%%%%%%%%%%%%%%%%%%%%%%%%%
In this letter we have investigated the influence of final-state
$NN$-rescattering effect on the linear photon asymmetry $\Sigma$ for
the $\vec\gamma d\to \pi^-pp$ reaction in the photon energy range from
$\pi$-threshold to $500$ MeV. We have found that the effect due to the
final two-nucleon interaction to be small, but it can has significant
contribution to the photon asymmetry through its interference with the
dominant term from the impulse approximation.  Furthermore, the linear
photon asymmetry is found to be sensitive to the energy of the
incident photon. In comparison with the preliminary experimental data
from LEGS \cite{LEGS}, the inclusion of $NN$-FSI
effect leads to a better agreement with experimental data.  With
respect to the comparison with the preliminary results of \cite{Lee}
in the IA, we found a large difference between both calculations in the
peak position. The origin of this difference is still not clear.

We would like to conclude that the results presented here for linear
photon asymmetry of $d(\vec\gamma,\pi^-)pp$ can be used as a basis for
the simulation of the behaviour of this asymmetry and for an optimal
planning of new experiments of this reaction with polarized photon
beams. An experimental check of these predictions for the linear
photon asymmetry covering a large range for the pion angle would provide
an additional significant test of our present understanding of this
spin observable. Furthermore, an independent evaluation in the
framework of effective field theory would be very interesting. Future
improvements should include further investigations including FSI as
well as two-body effects. 
%%%%%%%%%%%%%%%%%%%%%%%%%%%%%%%%%%%%%%%%%%%%%%%%%%%%%%%%%%
\begin{ack}
  This work is supported in part by the Bibliotheca Alexandrina
  -Center for Special Studies and Programs- under grant number:
  2602314 Sohag 2$^{nd}$-Sohag. I am indebted to Prof.\ H.\ Arenh\"ovel as
  well as the members of his work group for fruitful discussions and
  valuable comments.  I would like to thank Profs.\ T.-S.\ Harry Lee
  and T.\ Sato for useful discussions and Prof.\ S.A.E.\ Khallaf for a
  careful reading of this letter.
\end{ack}
%%%%%%%%%%%%%%%%%%%%%%%%%%%%%%%%%%%%%%%%%%%%%%%%%%%%%%%%%


\begin{thebibliography}{99}
  
\bibitem{Log00} A.Yu.\ Loginov, A.A.\ Sidorov, V.N.\ Stibunov, Phys.\ 
  Atom.\ Nucl.\ 63 (2000) 391.
  
\bibitem{Lee} T.-S.\ H.\ Lee, private communication; T.-S.\ H.\ Lee, in
  LQWq Workshop on Electromagnetic Nucleon Reactions at Low Momentum
  Transfer, August 23-25, 2001, Halifax, Nova Scotia, Canada.

\bibitem{Log04} A.Yu.\ Loginov, A.V.\ Osipov, A.A.\ Sidorov, V.N.\ 
  Stibunov, nucl-th/0407045.

\bibitem{Dar04a} E.M.\ Darwish, Nucl.\ Phys.\ A 735 (2004) 200.

\bibitem{Dar04b} E.M.\ Darwish, Int.\ J.\ Mod.\ Phys.\ E 13 (2004) 1191.

\bibitem{Dar04c} E.M.\ Darwish, J.\ Phys.\ G: Nucl.\ Part.\ Phys.\ G 31 
  (2005) 105.

\bibitem{Dar04d} E.M.\ Darwish, Nucl.\ Phys.\ A 748 (2005) 596.

\bibitem{Go02} St.\ Goertz, W.\ Meyer, G.\ Reicherz, Prog. Part. Nucl.
  Phys. 49 (2002) 403 [Erratum-{\it ibid} 51 (2003) 309].
  
\bibitem{GDH02} See, for example, Proceedings of the 2nd Int.\ 
  Symposium on {\it the Gerasimov-Drell-Hearn 2002 Sum Rule and the
    Spin Structure of the Nucleon}, Genova, Italy, 3-6 July, 2002,
  edited by M.\ Anghinolfi, M.\ Battaglieri, R.\ De Vita, (World
  Scientific, Singapore, 2003).
  
\bibitem{Kru03} B.\ Krusche, S.\ Schadmand, Prog.\ Part.\ Nucl.\ 
  Phys.\ 51 (2003) 399.
  
\bibitem{Bur04} V.\ Burkert and T.-S.\ H.\ Lee, Int.\ J.\ Mod.\ Phys.\ 
  E 13 (2004) 1035.

\bibitem{LEGS} A.\ Sandorfi, M.\ Lucas, private communication; M.\ 
  Lucas, in {\it ${\mathcal LOWq}$ 2001 Workshop on Electromagnetic
    Nuclear Reactions at Low Momentum Transfer}, Halifax, Nova Scotia,
  Canada, 23-25 August, 2001; A.\ Sandorfi, in \cite{GDH02}.

\bibitem{Pedroni} P.\ Pedroni, private communication; 
  C.A.\ Rovelli, Diploma Thesis, University of Pavia, Italy,
  (2002). 

\bibitem{Dar03} E.M.\ Darwish, H.\ Arenh\"ovel, M.\ Schwamb, Eur.\ 
  Phys.\ J.\ A 16 (2003) 111.

\bibitem{Lev01} M.I.\ Levchuk, M.\ Schumacher, F.\ Wissmann, 
  nucl-th/0011041.

\bibitem{Aren88}
        H.\ Arenh\"ovel, Few-Body Syst. 4 (1988) 55.

\bibitem{ScA96} R.\ Schmidt, H.\ Arenh\"ovel, P.\ Wilhelm, Z.\ Phys.\ 
  A 355 (1996) 421.
  
\bibitem{La+81} M.\ Lacombe {\it et al.}, Phys.\ Lett.\ B 101 (1981)
  139.
  
\bibitem{HaP8485} J.\ Haidenbauer, W.\ Plessas, Phys.\ Rev.\ C 30
  (1984) 1822; {\it ibid} C 32 (1985) 1424.
  
\bibitem{SL} T.\ Sato, T.-S.\ H.\ Lee, Phys.\ Rev.\ C 54 (1996) 2660.

\end{thebibliography}
\end{document}